\documentclass[twocolumn]{aastex63}
\usepackage[utf8]{inputenc}

\hypersetup{linkcolor=red,citecolor=blue,filecolor=cyan,urlcolor=magenta}

\usepackage[]{units}
\usepackage{gensymb}
\usepackage{amsmath}
\usepackage{booktabs}
\usepackage{romannum}
\usepackage{longtable}
\usepackage[final]{animate}

\newcommand{\sii}{\ion{S}{2}}
\newcommand{\nii}{\ion{N}{2}}

\newcommand{\oi}{\ion{O}{1}}
\newcommand{\oii}{\ion{O}{2}}
\newcommand{\oiii}{\ion{O}{3}}
\newcommand{\ha}{H\ensuremath{\alpha}}
\newcommand{\hb}{H\ensuremath{\beta}}

\graphicspath{{./}{figures_eps/}}

\shorttitle{SDSS MaNGA: Barred Milky Way Analogs}
\shortauthors{DK et al.}

\begin{document}

\title{The Effect of Bars on the Ionized ISM: Optical Emission Lines from Milky Way Analogs}
\correspondingauthor{Dhanesh Krishnarao}
\email{krishnarao@astro.wisc.du}

\author[0000-0002-7955-7359]{Dhanesh Krishnarao}
\affiliation{Department of Astronomy, University of Wisconsin-Madison, Madison, WI, USA}

\author[0000-0003-3097-5178]{Christy Tremonti}
\affiliation{Department of Astronomy, University of Wisconsin-Madison, Madison, WI, USA}

\author[0000-0001-9557-5648]{Amelia Fraser-McKelvie}
\affiliation{School of Physics and Astronomy, University of Nottingham, University Park, Nottingham, UK}

\author{Katarina Kraljic}
\affiliation{Institute for Astronomy, University of Edinburgh, Edinburgh, UK}

\author{Nicholas Fraser Boardman}
\affiliation{Department of Physics and Astronomy, University of Utah, Salt Lake City, UT, USA}

\author[0000-0003-0846-9578]{Karen L. Masters}
\affiliation{Department of Physics and Astronomy, Haverford College, Haverford, PA, USA}

\author[0000-0002-8109-2642]{Robert A. Benjamin}
\affiliation{Department of Physics, University of Wisconsin-Whitewater, Whitewater, WI, USA}

\author[0000-0002-9947-6396]{L. Matthew Haffner}
\affiliation{Physical Sciences Department, Embry-Riddle Aeronautical University, Daytona Beach, FL, USA}

\author{Amy Jones}
\affiliation{Department of Physics and Astronomy, University of Alabama, Tuscaloosa, AL, USA}

\author[0000-0003-4843-4185]{Zachary J. Pace}
\affiliation{Department of Astronomy, University of Wisconsin-Madison, Madison, WI, USA}

\author{Gail Zasowski}
\affiliation{Department of Physics \& Astronomy, University of Utah, Salt Lake City, UT, USA}

\author[0000-0002-3131-4374]{Matthew Bershady}
\affiliation{Department of Astronomy, University of Wisconsin-Madison, Madison, WI, USA}
\affiliation{South African Astronomical Observatory, Cape Town, South Africa}

\author[0000-0002-9808-3646]{Dmitry Bizyaev}
\affiliation{Apache Point Observatory, Sunspot, NM, USA}
\affiliation{Sternberg Astronomical Institute, Moscow State
University, Moscow, Russia}

\author{Jonathan Brinkmann}
\affiliation{Apache Point Observatory, Sunspot, NM, USA}

\author[0000-0002-8725-1069]{Joel R. Brownstein}
\affiliation{Department of Physics and Astronomy, University of Utah, Salt Lake City, UT, USA}

\author[0000-0002-7339-3170]{Niv Drory}
\affiliation{McDonald Observatory, The University of Texas at Austin, Austin, TX, USA}

\author[0000-0002-9808-3646]{Kaike Pan}
\affiliation{Apache Point Observatory, Sunspot, NM, USA}

\author[0000-0002-9808-3646]{Kai Zhang}
\affiliation{Lawrence Berkeley National Laboratory, Berkeley, CA, USA}

\begin{abstract}

Gas interior to the bar of the Milky Way has recently been shown as the closest example of a Low Ionization (Nuclear) Emission Region--LI(N)ER--in the universe. To better understand the nature of this gas, a sample of face-on galaxies with integral field spectroscopy are used to study the ionized gas conditions of 240 barred and 250 nonbarred galaxies, focusing on those that are most similar to the Milky Way. Strong optical line emission of [\nii] $\lambda 6584$, \ha, [\oiii] $\lambda 5007$, and \hb\ are used to diagnose the dominant ionization mechanisms of gas across galaxies and the Galaxy via Baldwin-Phillips-Terlevich (BPT) Diagrams. Barred galaxies show a strong suppression of star formation and an increase in composite and LI(N)ER like spectra in their inner regions when compared with similar nonbarred counterparts. This effect is lessened in galaxies of very low ($\log_{10}(\nicefrac{M_\star}{M_\odot}) \lesssim 10.4$) or very high ($\log_{10}(\nicefrac{M_\star}{M_\odot}) \gtrsim 11.1$) total stellar mass. Bar masks from Galaxy Zoo:3D show the bar's non-axisymmetric effect on the ionized gas and help predict the face-on distribution of ionized gas conditions near the bar of the Milky Way.

\end{abstract}


\section{Introduction}\label{sec:intro}
Optical line emission was first detected in an extragalactic system in 1909 \citep{Fath1909}, although its extragalactic nature was not posited until 1929 when Cepheids were observed in M31 \citep{Hubble1929a}. The correlation between distance and recession velocity \citep{Hubble1929b}---the Hubble-Lema\^itre law---sparked large spectroscopic campaigns of galaxy observations \citep[see][]{Humason1956}. In particular, a coordinated effort at the Mount Wilson and Lick Observatories started in 1935 and showed many galaxies with [\oii] $\lambda 3727$ \AA\ nebular emission in their central regions \citep{Humason1936, Mayall1939, Baade1951}. Around the same time, \citet{Babcock1939} tentatively detected [\oii] $\lambda 3727$ \AA\ emission from the inner $\sim 1$ kpc of M31 \citep[later confirmed by][]{Munch1960}, where no star formation is present. Samples of galaxies showing strong optical emission lines from ionized gas continued to rise and soon \citet{Seyfert1943} discovered bright highly ionized gas in the centers of some galaxies. 

\citet{Burbidge1962,Burbidge1965} confirmed and extended these initial ionized gas detections using \ha\ and [\nii] $\lambda 6584$ \AA\ emission across $85$ galaxies and found the line ratio of [\nii]/\ha\ increased from $~\nicefrac{1}{3}$ in outer regions and spiral arms to at or above unity in many galactic nuclei. Ratios this large were not seen within the Milky Way both at that time \citep{Courtes1960} and for the next half century \citep{Haffner2009}. \citet{Burbidge1965} noted that such large line ratios could only be explained with an enhanced nitrogen abundance or a large electron temperature of $10000 - 20000$ K \citep{Burbidge1963}. Over time, additional optical emission lines were studied and larger samples of galaxies were categorized based on their line ratios and underlying sources of ionization \citep{Heckman1980, Baldwin1981}. 

Baldwin-Phillips-Terlevich \citep[BPT;][]{Baldwin1981} diagrams use a combination of the [\nii]/\ha\ and [\oiii]/\hb\ line ratio to classify extragalactic systems. This method can identify the dominant source of excitation in galaxies using only line emission, allowing for source classification in large surveys. \citet{Veilleux1987} improved on this method and many theoretical and empirical studies of large galaxy samples yielded four broad categories based on the source of ionization: star formation, active galactic nuclei (AGN), composites, and Low Ionization Nuclear Emission Regions---LINERs \citep{Kewley2001, Kauffmann2003, Schawinski2007}. While star formation and AGN have known sources of ionization, LINERs are a descriptive name for an unknown source first coined by \citet{Heckman1980}. In this work, we use the ``nuclear" term in parentheses (LI(N)ER) since in many cases, this class of emission is seen outside of galactic nuclei \citep[e.g.][]{Sharp2010, Annibali2010,Belfiore2016}. Some LI(N)ERs may arise from low luminosity AGN, while others are correlated with older stellar populations \citep{Kewley2006, Belfiore2016}. The characteristic signatures of LI(N)ERs are lower values for [\oiii]/\hb\ than in AGN, but higher ratios for lower ionization lines, such as [\nii]/\ha, [\sii]/\ha, and [\oi]/\ha\ \citep{Ho1993, Ho1997, Filippenko2003}. The large [\nii]/\ha\ line ratios require a significant source of ionizing radiation in regions of galaxies where little to no star formation is occurring as \citet{Babcock1939} observed for M31.

Until recently, LI(N)ER gas was only observable in extragalactic systems, with the closest such example being M31. Optical observations of the ionized gas counterpart to the neutral Milky Way structure, the Tilted Disk \citep{BL1, BL2, BL3}, have allowed for the inner Milky Way to be classified using the same diagnostic method of BPT diagrams \citep{Krishnarao2020a}. The Milky Way has been shown to have a bar \citep[cf.][]{JBH2016} most recently as a stellar density enhancement using Gaia, Pan-STARRS1, 2MASS, and AllWISE data \citep{Anders2019}. Gas interior to the bar around $0.5 - 1.5$ kpc from Galactic Center is largely ionized with optical emission line ratios similar to LI(N)ERs \citep{Krishnarao2020a}. This gas is now the closest LI(N)ER to us in the universe at a distance of only $\sim 7$ kpc. While the rest of Milky Way does not yet have widespread measurements of all four emission lines used for the BPT diagram, it is rare locally and in spiral arms to have [\nii]/\ha\ line ratios as large as is seen in LI(N)ERs \citep[see][for a review]{Haffner2009}. This suggests that LI(N)ER gas in the Milky Way is constrained to the vicinity of the bar and raises the question of how important a bar is in setting the ionized gas conditions and radiation field.

Simulations of gas in barred galaxies predict strong shear and non-circular flows across many phases of gas \citep{Renaud2013, Seo2019, Sormani2019}. They also predict concentrated locations of star formation in the form of rings along with bursty star formation in a Central Molecular Zone-like core. Such ring-like structures have been seen in optical images \citep[e.g.][]{Buta1986} and are starting to be studied in nearby barred galaxy observations with the MUSE IFU \citep[see][and the TIMER survey]{Gadotti2019}. Galaxies with redder colors and stronger bulges have been shown to have larger bar fractions \citep{Masters2011}. Stars in the bar also show flatter age and metallicity gradients than stars in the disk at similar radii \citep{Fraser-McKelvie2019}, suggesting a large degree of radial mixing in the bar. Barred galaxies also have a lower atomic gas content, potentially inhibiting star formation in their host galaxy \citep{Masters2012, Newnham2019}. In addition, \citet{Hoyle2011} found that longer bars are redder and are present in redder galaxies, establishing a strong link between star formation suppression and bars. \citet{George2019} showed that in M95, star formation quenching in the vicinity of the bar is consistent with strong gas inflows that leave the bar devoid of neutral and molecular hydrogen that can form stars. However, even with this suppression of star formation, atomic ionized gas is still present and often displays enhanced optical line ratios similar to LI(N)ERs. 

Since the bar provides a non-axisymmetric potential, gas is expected to transition from circular rotation to non-circular rotation near the bar radius. The effect of the resulting shocks associated with this transition and gas flows in a bar potential should be seen as changes in the BPT classifications of gas within galaxies. Kinematic models are required to analyze the distribution of gas in the inner Milky Way, where radial velocities must be mapped to distances. While some models of this nature are available, such as with the Tilted Disk in the inner Milky Way \citep{Krishnarao2020a}, a complete picture cannot be seen from within the Galaxy. In particular, the behavior of the ISM interior to the bar radius but outside of the bar itself is difficult to ascertain with direct observations in our Galaxy. Face-on extragalactic systems provide a better way to obtain a more complete view of the ISM in the vicinity of a bar, both interior to and outside of its non-axisymmetric distribution. 

Studies focusing on extragalctic Milky Way analogs help place the Milky Way on extragalactic scaling relations \citep[e.g.][]{Robotham2012, Geha2017, Boardman2019, Fraser-McKelvie2019}.
These analog samples have been used to infer statistical properties of Milky Way-like galaxies and explore the extent such systems are outliers from global galaxy relations \citep{Licquia2015, LicquiaAndNewman2015, Boardman2019}. They are selected based on stellar mass, star formation rate, bulge prominence, or other global properties, but not on the presence of finer morphological structures, such as a bar. Since a bar can have large scale impacts on the interstellar medium (ISM) and star formation, it could be a vital characteristic of a galaxy to consider when constructing a sample of Milky Way analogs. In this work, we focus on better understanding LI(N)ERs and determining how crucial the presence of a bar is for the ionized gas properties of galaxies using SDSS MaNGA. 

This paper is organized as follows. In Section \ref{sec:obs}, we describe the SDSS MaNGA observations and additional data products used and explain our sample selection process in Section \ref{sec:sample}. Section \ref{sec:stats} presents our comparisons of barred and nonbarred galaxies as a function of various galaxy properties and Section \ref{sec:bar_specific} focuses on the barred sample to reveal the non-axisymmetric effects on the ISM of galaxies. We discuss the broader implications of our findings in Section \ref{sec:disc} in the context of the Milky Way and close with our conclusions in Section \ref{sec:conclude}. 

\section{Data}\label{sec:obs}

This paper uses data and products associated with the Sloan Digital Sky Survey (SDSS) \citep{York2000}. The details of these data products are described below. 

\subsection{MaNGA IFU Data}
The Mapping Nearby Galaxies at Apache Point Observatory (MaNGA) survey is a part of SDSS-\Romannum{4} \citep{Blanton2017} that uses integral field spectroscopy to observe $3D$ spectra of nearly $10,000$ galaxies \citep{Bundy2015}. Spectra are obtained with the BOSS spectrograph on the 2.5 meter telescope at Apache Point Observatory \citep{Gunn2006} at a spectral resolution of $R \sim 2000$ for $\unit[3600]{\text{\AA}} < \lambda < \unit[10300]{\text{\AA}}$ and have exposure times such that a target signal-to-noise (S/N) level is achieved \citep{Bundy2015}. Fibers subtend $\unit[2]{"}$ on the sky \citep{Smee2013} and are bundled into integral field units (IFUs) with sizes ranging $\unit[12]{"} - \unit[32]{"}$ in diameter with $19 - 127$ fibers \citep{Drory2015}.
Observed data undergo sky subtraction and flux calibration using simultaneous observations of the sky and standard stars \citep{Yan2016}. The median point spread function for MaNGA data cubes is $\unit[2.5]{"}$ and roughly corresponds to kiloparsec physical scales at the targeted redshift range ($0.01< z < 0.15$). Observations are dithered and mapped onto spectroscopic pixels (or spaxels) that are $\unit[0.5]{"}$ across.

MaNGA targets local galaxies and the sample is selected to have a flat distribution of i-band absolute magnitude and uniform radial coverage. This sample is composed of three main components, a primary sample where $80\%$ of galaxies are covered out to $1.5\ R_e$, a secondary sample where $80\%$ of galaxies are covered out to $2.5\ R_e$, and a color enhanced supplement to improve coverage of poorly sampled regions of the $NUV-i$ vs. $M_i$ color-magnitude plane \citep{Wake2017}. 
All MaNGA data in this work are reduced using \texttt{v2\_5\_3} of the MaNGA Data Reduction Pipeline \citep[DRP;][]{Law2016} and employs the Data Analysis Pipeline \citep[DAP;][]{Westfall2019,Belfiore2019} from the internal eighth MaNGA product launch (MPL-8) that contains observations of $6507$ galaxies.  The DAP provides emission line data products that this work employs ([\nii] $\lambda 6584$, \ha, [\oiii] $\lambda 5007$, and \hb) and can be accessed using the python package \href{https://sdss-marvin.readthedocs.io/en/latest/index.html}{\texttt{sdss-marvin}} \citep{Cherinka2019}. 
The processed data files used here follow a hybrid binning scheme where spectra are first Voronoi binned \citep{Cappellari2003} to reach a g-band S/N $\sim 10$ for determining stellar kinematics, while emission lines are fit to each spaxel within the bin. Throughout this work, we assume a standard cosmology of WMAP9 \citep{Hinshaw2013}.

\subsection{Galaxy Zoo 2}
Morphological classifications for all SDSS MaNGA parent sample galaxies are from the citizen science Galaxy Zoo 2 project \citep{GalaxyZoo2, Hart2016}. Volunteer citizen scientists used SDSS galaxy images to identify galaxies as early-types, late-types, or mergers, along with measuring finer features such as bars, bulges, and shapes of edge-on disks. Classifications are made following a decision tree model, where users answer questions about a galaxy and receive follow-up questions based on their response. \citet{GalaxyZoo2} provides unbiased vote fractions, improved upon in \citet{Hart2016}, for each of these questions along with thresholds to determine well sampled galaxies for each classification. In this work, the thresholds recommended in \citet[][their Table 3]{GalaxyZoo2} are used to find a barred and nonbarred galaxy sample.

\subsection{Total Stellar Masses and Star Formation Rates}
Total stellar masses for MaNGA galaxies are from the NASA-Sloan Atlas \citep[NSA;][]{Blanton2011} and derived from an elliptical Petrosian photometric analysis. Estimates of the total star formation rate (SFR) are from the DAP \citep{Westfall2019, Belfiore2019} and are calculated using the Gaussian fitted \ha\ flux within the IFU field-of-view following the work of \citet{Kennicutt2012, Murphy2011, Hao2011} and assuming a Kroupa Initial Mass Function \citep[IMF;][]{Kroupa2001}. These SFR measurements do not remove non-star forming spaxels(e.g. AGN or LI(N)ER BPT classifications) or correct for extinction. They provide an estimate of the star formation within the IFU footprint of a galaxy and are only used as a brief check. 

\subsection{Stellar Mass Surface Densities}
\citet{Pace2019a} provide robust estimates of resolved stellar mass for all SDSS MaNGA galaxies from MPL-8 using a principal component analysis (PCA) method, for a more computationally-tractable stellar continuum fit. A set of 6 eigenspectra is defined by performing PCA on a set of 40,000 synthetic ``training" spectra, themselves obtained from applying procedurally-generated star-formation histories to a theoretical library of stellar atmospheres \citep[see][]{Conroy2012} using the \citet{Kroupa2001} IMF. 
The PCA basis set of six eigenspectra span a lower-dimensional space inside which the spectral-fitting occurs, but a space which expresses $>99\%$ of the variance in the training library. Each spectrum (whether from the training set or observed) can be expressed as a linear combination of the six eigenspectra:
\begin{equation}
    S_{\lambda} = Q \left( S_{med} + \sum_i A_{PC,i} \cdot S_{PC,i} \right)
\end{equation}
where $Q$ is a normalization factor, $S_{med}$ is the median basis spectrum, $A_{PC,i}$ are the amplitudes of the principal component spectra, and $S_{PC,i}$ are the principal component eigenspectra. The SDSS i-band stellar mass-to-light ratio for a given spectrum $S_{\lambda}$ is obtained by evaluating the likelihood of each training spectrum $S_{\lambda-trn}$ against the observation $S_{\lambda}$ in the lower-dimensional principal component space. The corresponding values of stellar mass-to-light ratio for each $S_{\lambda-trn}$ are then treated like samples from a prior, and are weighted according to their likelihood: that weighted sample median is taken as the best estimate of stellar mass-to-light ratio. This method is reliable at a wide range of signal-to-noise ratios $(2 < \text{S/N} < 30)$, at low-to-moderate dust attenuation $(\tau \lesssim 4)$, and over the full range of realistic stellar metallicities.

The stellar mass-to-light ratios are converted to stellar mass surface density, first by computing and multiplying by the reconstructed i-band magnitude of the starlight from the principal component fit, then dividing by the projected area of a MaNGA spaxel at the galaxy's redshift, and finally deprojecting using the minor-to-major axis ratio from the NSA \citep{Blanton2011}.

\subsection{Galaxy Zoo:3D}
Ongoing citizen science efforts associated with the Galaxy Zoo project provide masks to isolate emission from structural components of galaxies, including bars. These Galaxy Zoo:3D \citep{Masters_inprep} masks allow for emission originating within the bar to be compared with emission outside of the bar. Details of this ongoing project can be found on their website \href{https://www.zooniverse.org/projects/klmasters/galaxy-zoo-3d}{(https://www.zooniverse.org/projects/klmasters/galaxy-zoo-3d)}, and preliminary data products have been used to study density wave theory \citep{Peterken2019} and stellar population gradients \citep{Fraser-McKelvie2019}. While qualitative in nature, these masks effectively separate MaNGA spaxels dominated by bar light from the rest of the galaxy. Spaxels flagged to be in the bar by at least $20\%$ of participants are considered to be in the bar, as done in \citet{Fraser-McKelvie2019}. Out of the $240$ barred galaxies in our full MPL-8 sample (described in Section \ref{sec:sample}), $213$ have reliable bar masks. The maximum radius of spaxels in a bar mask and the minimum radius of spaxels outside a bar mask provide measures of the length and width of the bar, respectively. 
These bar length measurements are validated with independent measures of the bar length using a Fourier decomposition method employed in \citet[][see Appendix \ref{sec:bar_valid} for details]{Kraljic2012}.

\section{Galaxy Sample}\label{sec:sample}
The ISM is known to be strongly affected by bars in galaxies as evidenced by many N-body simulations \citep[e.g][]{Athanassoula1992, Debattista1998} and observational studies \citep[e.g.][]{Hoyle2011, Masters2012, George2019}.
Because the Milky Way is a prominently barred galaxy \citep{Binney1991,Weiland1994, Benjamin2005,JBH2016}, it is crucial to understand the impact of bars on the ISM, but doing so from within the Galaxy is limited by our view from within. Instead, face-on galaxies allow us to examine varying ionized gas properties of the ISM as a result of the presence of bars. Four samples of galaxies are considered: barred Milky Way analogs, nonbarred analogs, a full barred galaxy sample, and a full nonbarred galaxy sample. The analog samples are constructed of galaxies in the MaNGA parent sample based on morphology, color, and stellar mass, while our full barred and nonbarred samples follow the same criteria but do not consider a stellar mass range. A mass selection is applied to nonbarred galaxies to match the stellar masses of barred galaxies. 

\begin{figure}[ht!]
\epsscale{1.15}
\plotone{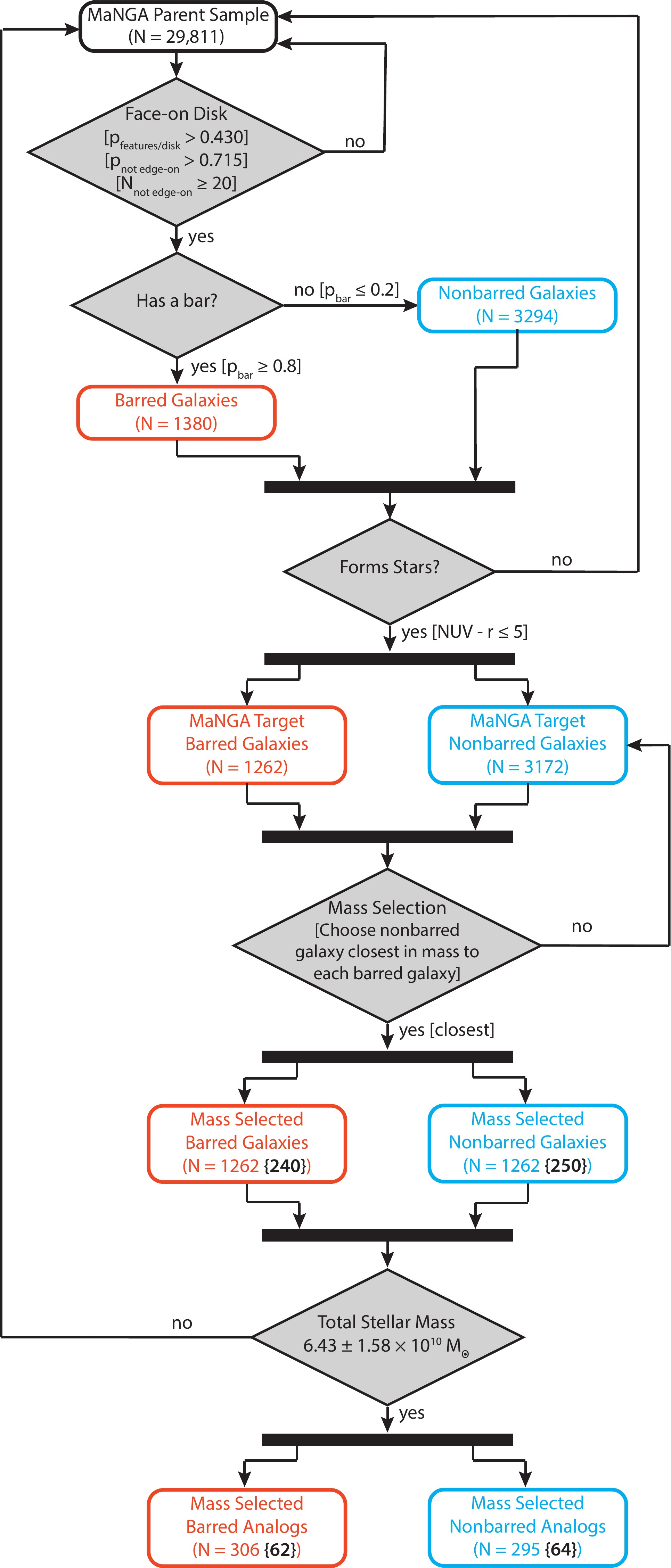}
\caption{Sample Selection criteria used as a decision tree. Decision making criteria are shown in gray diamonds, including specific Galaxy Zoo cutoff parameters \citep{GalaxyZoo2}, color \citep[as done in][]{Belfiore2017}, and stellar mass \citep{McMillan2011}. Sample sizes ($N$) are shown along each step and the MPL-8 observed sample used in this work are shown in curly brackets with boldface black numbers.  \label{fig:SampleTree}}
\end{figure}

The complete MaNGA survey will observe approximately $10,000$ out of $29,811$ galaxies from the parent sample. Our sample is initially constructed from all galaxies in this parent sample. This allows for our results to be easily replicated with additional galaxies as they are observed up until the completion of the MaNGA survey. A decision tree of the sample selection process is shown in Figure \ref{fig:SampleTree} and described in the following subsections. 

\subsection{Morphology}
Firstly, the morphology of the SDSS MaNGA parent sample of galaxies are restricted using Galaxy Zoo 2 \citep{GalaxyZoo2} to those with a face-on disk. This process starts with selecting galaxies with at least $43\%$ of votes identifying a disk ($p_{features/disk} > 0.43$), followed by at least $71.5\%$ of votes identifying the galaxy as not edge-on ($p_{not\ edge-on} > 0.715$) with at least $20$ such votes ($N_{not\ edge-on} \geq 20$). These face-on disk galaxies are then sorted into galaxies with and without a bar using at least $80\%$ of votes identifying a bar ($p_{bar} \geq 0.8$; barred galaxies) and less than $20\%$ of votes identifying a bar ($p_{bar} \leq 0.2$; nonbarred galaxies). This results in a total of $1380$ barred galaxies and $3294$ nonbarred galaxies in the SDSS MaNGA parent sample. These parameters are chosen based on the recommendations of \citet{GalaxyZoo2}, and the $p_{bar} \geq 0.8$ parameter is a very tight constraint on the initial set of galaxies. This tight constraint reduces our completeness of barred galaxies to about $29.4\%$ when compared with the visual classifications of \citet{Nair2010}, but ensures a high accuracy in identifying true bar structures. 

\subsection{Star Formation and Mass Selection}
\begin{figure}[ht!]
\epsscale{1.15}
\plotone{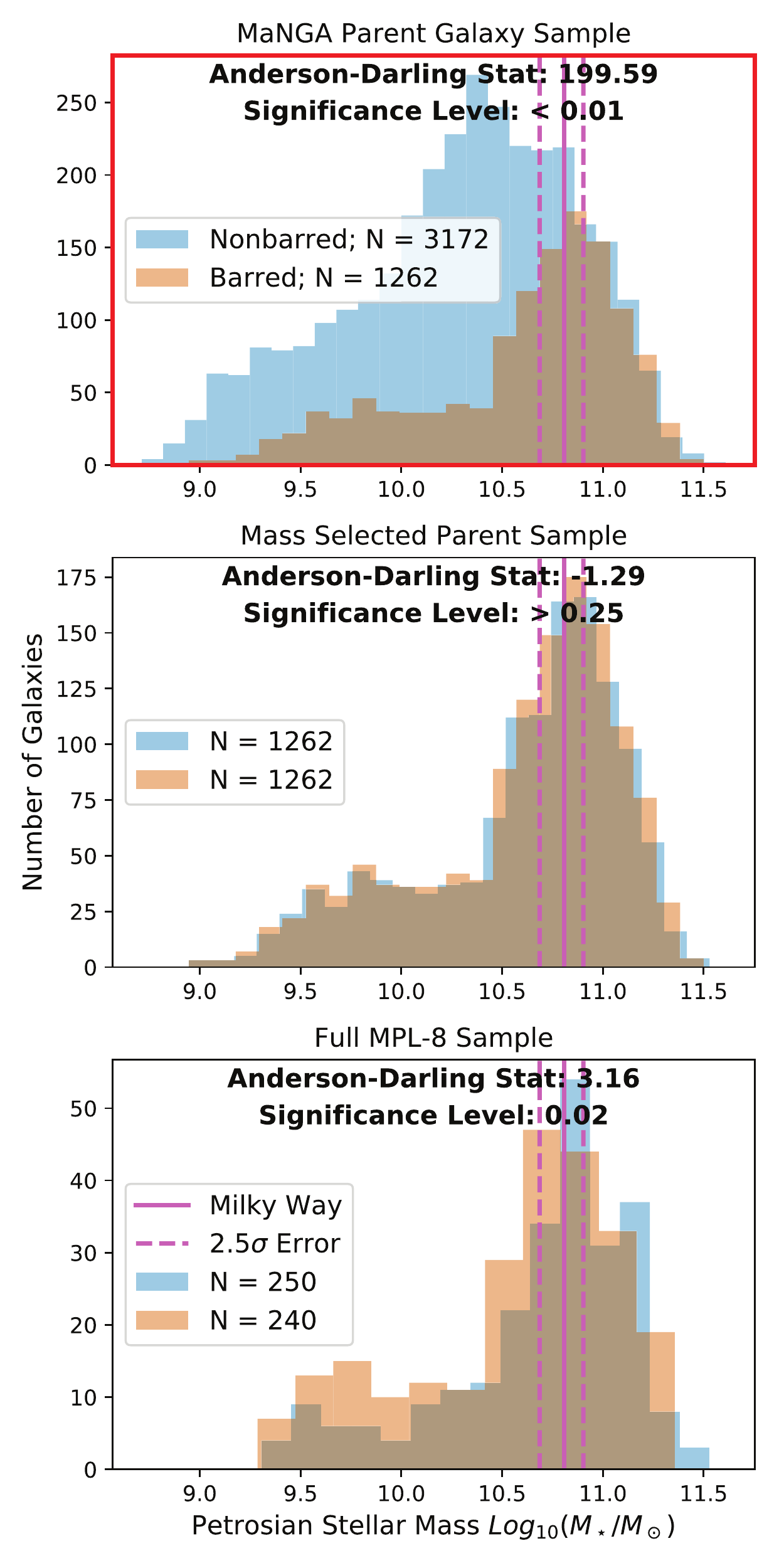}
\caption{Total stellar mass histograms for the entire SDSS MaNGA parent sample (top), our mass selected parent sample (middle) and the full MPL-8 observed sample of barred (orange) and nonbarred (blue) galaxies. The pink solid and dashed lines are for the Milky Way stellar mass and $2.5 \sigma$ uncertainties from \citet{McMillan2011}. Results of the Anderson-Darling test statistic and significance level that both samples are drawn from the same population are shown along the top with a red frame around those in which the null hypothesis can be rejected at the $p = 0.01$ level. \label{fig:massSelection}}
\end{figure}

To consider the ISM within our sample, we select galaxies that show signs of star formation using an $NUV - r \leq 5$ color cut as done in \citet{Belfiore2017}. We use this minimal cut rather than selecting galaxies based on star formation rate because of the difficulty in estimating an accurate star formation rate for all the MaNGA galaxies and, similarly, the difficulty in estimating a global star formation rate of the Milky Way. DAP estimates of the star formation rate do not account for extinction and take a simple approach to provide consistent measurements for all galaxies \citep{Westfall2019, Belfiore2019}. Within the Milky Way, systematic differences between different methods are still present \citep[see e.g.][]{McKee1997, Chomiuk2011}. This loose selection restricts our barred and nonbarred galaxies to $1262$ and $3172$, respectively. With these criteria, the barred and nonbarred galaxy samples show a noticeable difference in their total elliptical Petrosian stellar mass distributions from NSA \citep{Blanton2011}, with nonbarred galaxies showing a greater presence at lower masses. To ensure our sample selection process has not inadvertently selected on other parameters, we use the k-sample Anderson-Darling statistical test to compare many parameter distributions from our samples. 
The k-sample Anderson-Darling statistical test \citep{k-anderson-darling} compares the underlying populations for each sample with a null hypothesis that all samples come from the same population. We prefer the Anderson-Darling test to the more common Kolmogorov-Smirnov test for its more sensitive statistical properties \citep[see][]{Babu2006}. Ideally, the barred and nonbarred samples would fail to reject this null hypothesis for all galaxy parameters, indicating that our barred and nonbarred samples are probing the same population of galaxies. This ensures our study is not affected by other galaxy parameters that can impact the ISM, isolating the effects of the bar.

\begin{figure}[ht!]
\epsscale{1.2}
\plotone{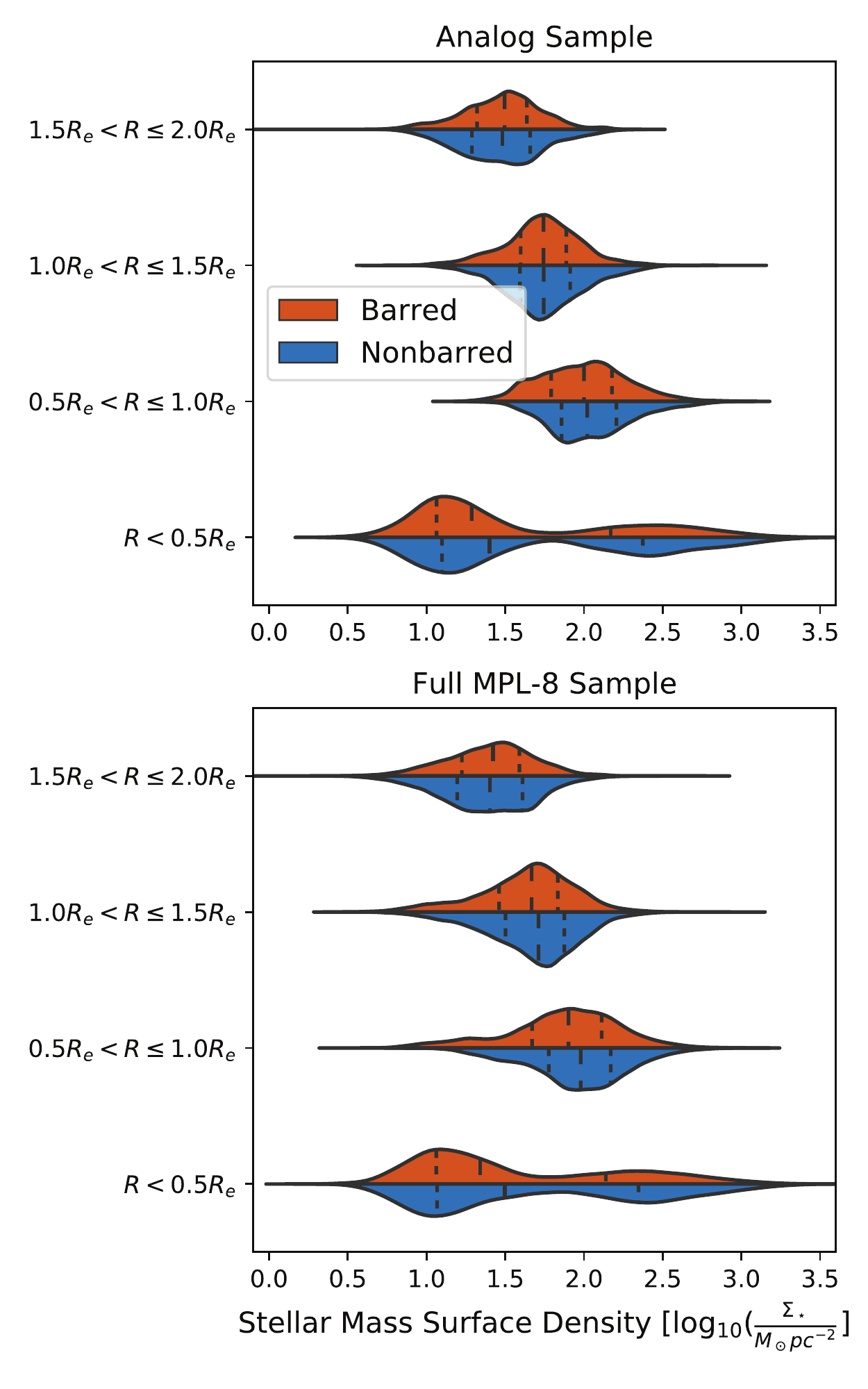}
\caption{Violin style Gaussian kernel density plots of the stellar mass surface density at different radial bins of barred (orange) and nonbarred (blue) galaxies normalized by their effective radius for analog (upper) and full MPL-8 (lower) samples. The interquartile range of each distribution is shown with dashed lines. The shapes of the violin plots are drawn based on the probability density distribution and have their widths normalized based on the number of spaxels in each category. The relative widths of the violin shapes across different categories at a given stellar mass surface density bin correspond to the fraction of spaxels within a category at that bin value. In general, both barred and nonbarred galaxies show the same distributions and trends.  \label{fig:PCA_Sigma_Re}}
\end{figure}

The stellar mass distributions for the parent barred and nonbarred samples allow for this null hypothesis to be rejected at the $0.01$ level, suggesting the barred and nonbarred sample come from distinct underlying populations. To avoid this stellar mass bias, a mass selection is applied to the nonbarred galaxies, where for each barred galaxy, the nonbarred galaxy closest in total stellar mass is chosen. Figure \ref{fig:massSelection} shows the results of this process and our mass selected barred and nonbarred samples show a statistically similar distribution, with the k-sample Anderson-Darling statistical test p-value $> 0.25$. Similar tests comparing the star formation rate \citep[from DAP][]{Westfall2019}, FUV, NUV, and i-band photometry \citep[from NSA][]{Blanton2011} of these samples also fail to reject the null hypothesis at the $p = 0.01$ level, indicating that the samples are drawn from similar populations. In this work, we use observations available as of MPL-8, which results in our full MPL-8 barred and nonbarred galaxy samples with sizes of $N = 240$ and $N = 250$, respectively. The bottom panel of Figure \ref{fig:massSelection} shows the total stellar mass distributions and statistical test results for the full MPL-8 samples. While small differences are present in these distributions, they are mostly similar (p-value $> 0.01$) and should provide a fair comparison of barred vs. nonbarred galaxies. 

The presence of a different underlying \textit{distribution} of stellar mass across our barred and nonbarred samples may introduce another source of bias. Differences in the ISM of our galaxy samples may be influenced by differences in the bulge properties of our two samples. To test this, we consider the stellar mass surface density across each galaxy for our barred and nonbarred samples. 
In Figure \ref{fig:PCA_Sigma_Re}, a violin style Gaussian kernel density histogram of the stellar mass surface density per spaxel from the PCA analysis of \citet{Pace2019a} is shown along different effective radius bins for barred and nonbarred galaxies. This style of plot is used throughout this paper to compare the distributions of properties across categories. The shapes of the violin plots are from the probability density distribution, where widths are normalized by the number of spaxels in each category. The relative widths of the violin shapes across different categories at a given bin correspond to the fraction of spaxels within a category at that bin value. While small differences can be seen in the inner $1\ R_e$, in general, both samples show similar radial distributions of their stellar mass. 

Lastly, all of the barred and nonbarred galaxy samples are composed of galaxies from approximately $50\%$ primary, $25\%$ secondary, and $25\%$ color enhanced samples \citep[see][]{Wake2017}, resulting in an equal sampling of the radial extent of galaxies. These tests confirm that differences seen between our barred and nonbarred galaxies should primarily result from the impact of the bar.

\subsection{Milky Way Analogs}
\begin{figure*}[ht!]
\epsscale{1.15}
\plotone{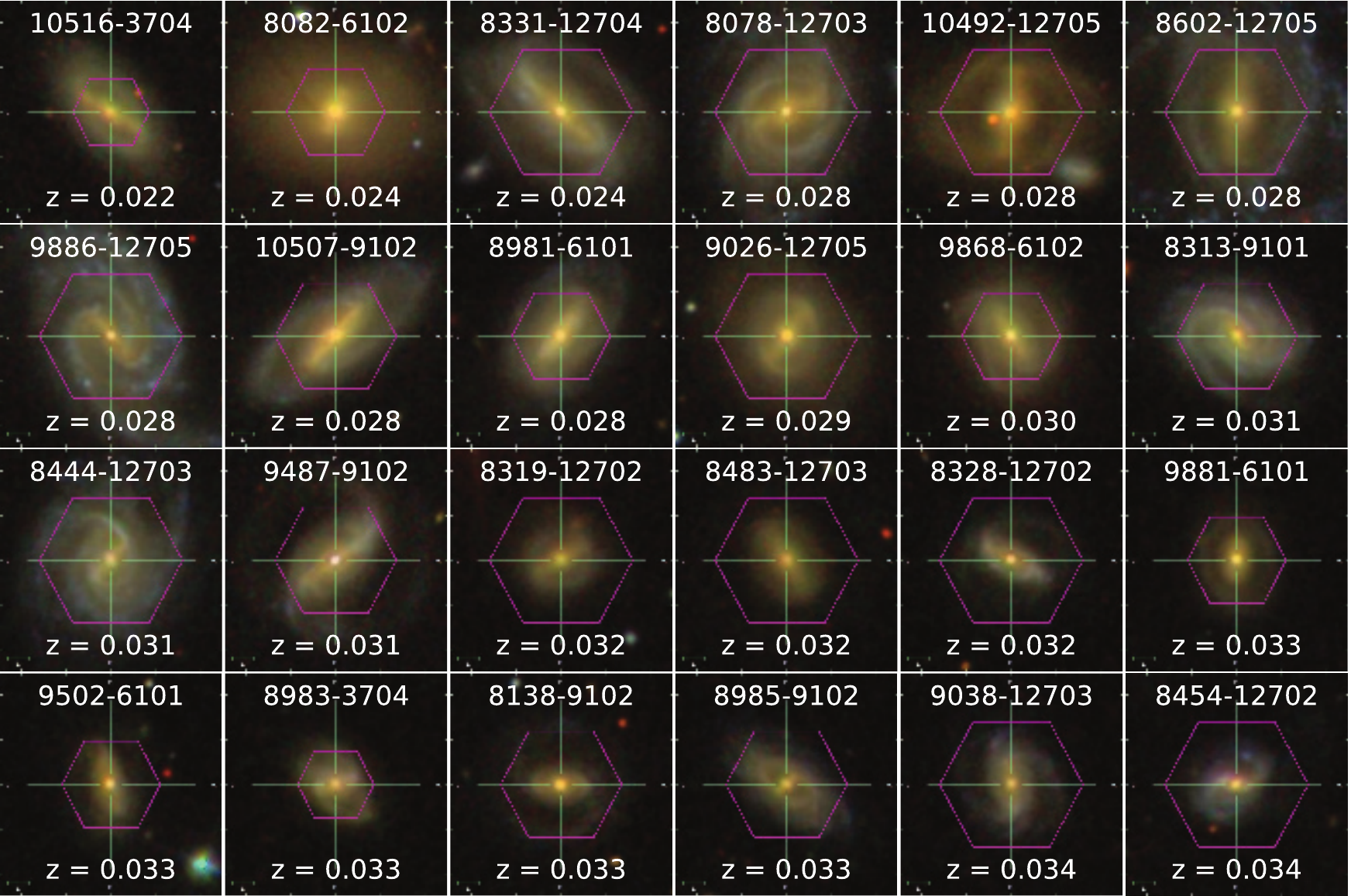}
\caption{SDSS images of $24/62$ barred Milky Way analogs that lie within $1\sigma$ of the Milky Way stellar mass estimate of \citet{McMillan2011} with their plate-ifu identifiers and redshift in white. The pink hexagon overlays the spatial extent of the MaNGA integral field unit. \label{fig:images}}
\end{figure*}

We apply a tight constraint on stellar mass based on the estimate of \citet{McMillan2011} for the Milky Way of
$
M_{\star} = 6.43 \pm 0.63 \times 10^{10}\mathrm{M_\odot}
$. Systematic errors in the NSA stellar mass estimates are known to reach up to $0.2$ dex \citep{Pace2019b}, so we loosen this constraint to encompass a range of $\pm 2.5 \sigma$ of the \citet{McMillan2011} estimate. This results in two ``Milky Way analog" samples with and without a bar of sample size $N = 62$ and $N = 64$, respectively.  

SDSS images of the $24$ barred analogs within $1 \sigma$ of the stellar mass criterion are shown in Figure \ref{fig:images}. In this work, we will focus on the differences between the barred and nonbarred ``Milky Way analog" samples and demonstrate the importance of a bar in defining analogs when considering the ISM. The larger full MPL-8 samples of barred and nonbarred galaxies will show the overall impact of a bar on the ISM across many galaxy scales. While more recent estimates of the total stellar mass of the Milky Way have shown lower values \citep[e.g.][]{Bovy2013, McMillan2017}, we do not expect our selection of a specific mass range to significantly affect our results. All masses are considered throughout our analysis to help understand the importance of the total stellar mass in defining analogs.

\section{Barred vs. Nonbarred Galaxies}\label{sec:stats}

\begin{figure}[ht!]
\epsscale{1.15}
\plotone{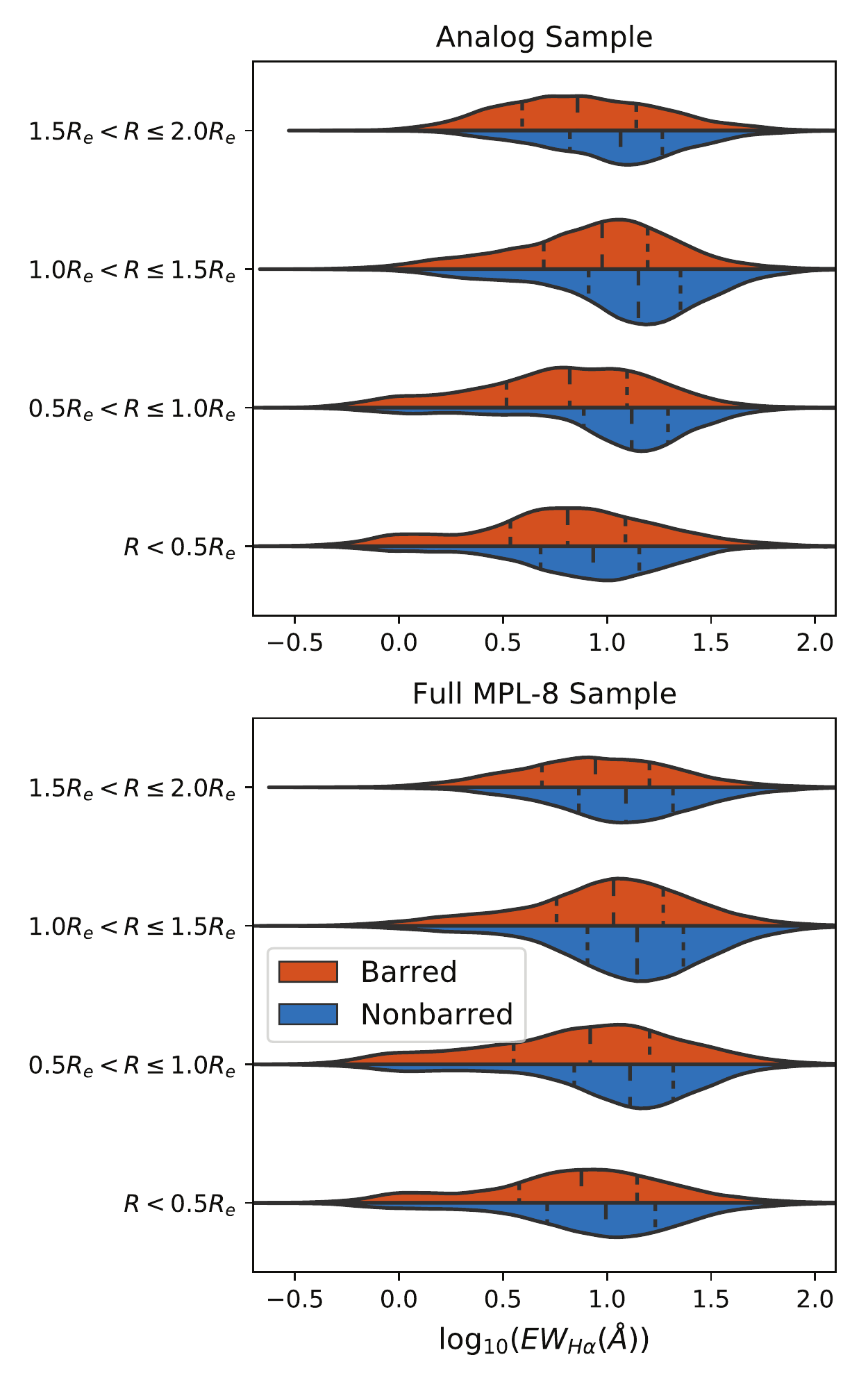}
\caption{Violin style Gaussian kernel density plots of the \ha\ equivalent width at different radial bins of barred (orange) and nonbarred (blue) galaxies normalized by their effective radius for analog (upper) and full MPL-8 (lower) samples. The interquartile range of each distribution is shown with dashed lines. The shapes of the violin plots are drawn based on the probability density distribution and have their widths normalized based on the number of spaxels in each category. The relative widths of the violin shapes across different categories at a given \ha\ equivalent width bin correspond to the fraction of spaxels within a category at that bin value. Barred galaxies show lower equivalent widths than nonbarred galaxies, especially at $0.5\ R_e < R \leq 1.0\ R_e$, suggesting that barred galaxies have less ionized gas in their inner regions than nonbarred galaxies. \label{fig:EW_Ha}}
\end{figure}

\begin{figure}[ht!]
\epsscale{1.15}
\plotone{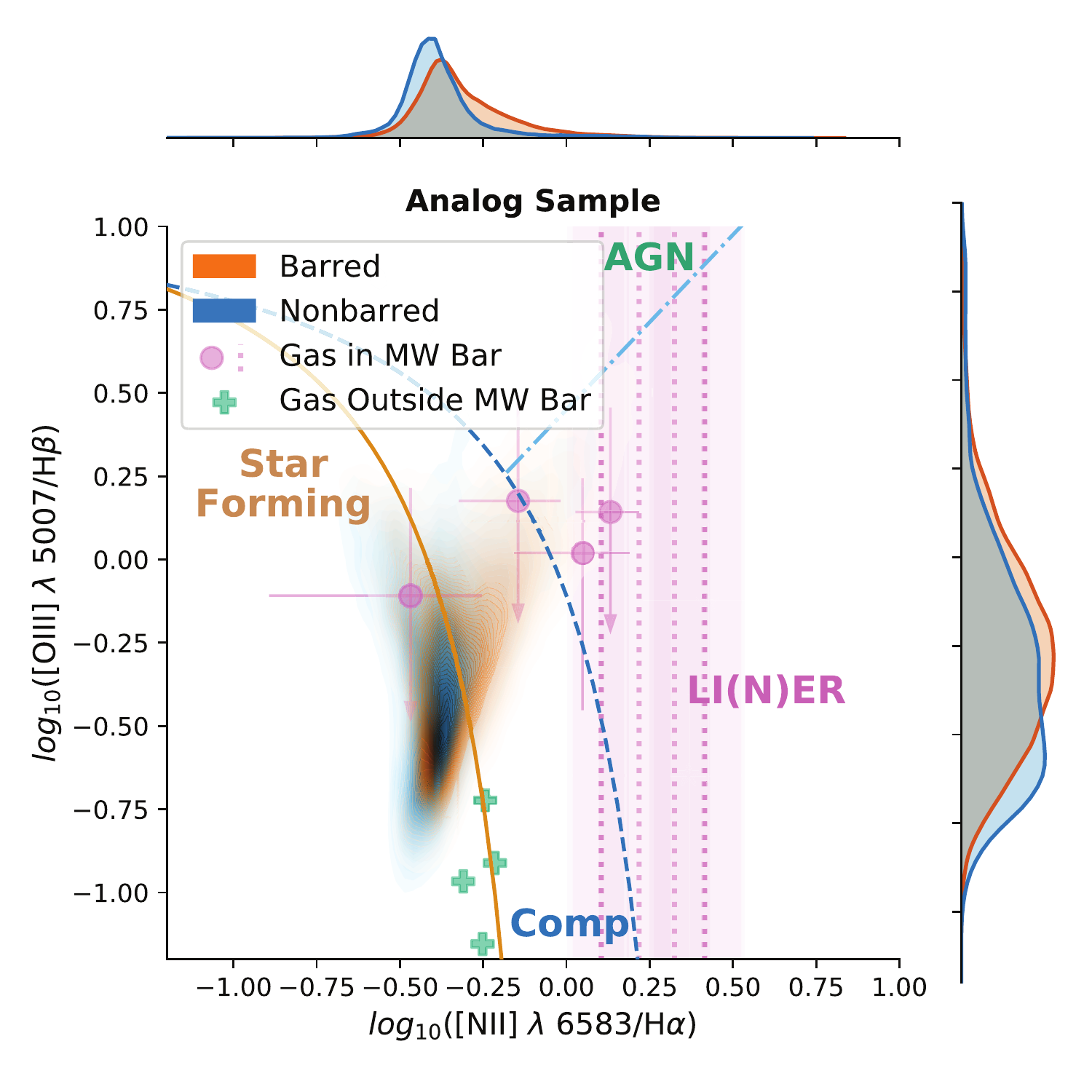}
\caption{[\nii]/\ha~Gaussian kernel density estimate BPT-diagram of the barred (orange) and nonbarred (blue) analog galaxy samples with Milky Way observations with $1$ $\sigma$ errors from \citet{Krishnarao2020a} in points and bars (pink) and from \citet{Madsen2005} in plus signs (green). The bars indicate cases where [\oiii]/\hb\ could not be constrained. Barred galaxies show a larger spread in [\nii]/\ha\ line ratios than their nonbarred counterparts. Classification lines are from \citep[][orange solid line]{Kauffmann2003}, \citep[][blue dashed line]{Kewley2001}, and \citep[][cyan dot-dashed line]{Schawinski2007}. \label{fig:BPT_KDE_MW}}
\end{figure}

Our barred and nonbarred galaxy samples provide a unique testing ground to study the influence of the bar on the ISM. Figure \ref{fig:EW_Ha} shows the distribution of the equivalent width of \ha\ at different radial bins for barred and nonbarred galaxies, revealing that barred galaxies tend to have less ionized gas than nonbarred galaxies in their inner regions. The ionization mechanisms of the gas in both samples are analyzed using BPT-diagrams. Figure \ref{fig:BPT_KDE_MW} shows barred and nonbarred galaxies on the [\nii]/\ha\ BPT diagram with Wisconsin H-Alpha Mapper (WHAM) Milky Way observations overlaid \citep{Krishnarao2020a, wham-nss}. MaNGA data are restricted to S/N $> 3$, which \citet{Belfiore2019} showed to be robust with minimal systematic errors for the emission lines of interest. To first order, barred galaxies show a larger spread in [\nii]/\ha\ line ratios than their nonbarred counterparts. In the following sections, this relationship is further explored, and we consider radial trends in the ionized gas properties across different total stellar masses. Because only the [\nii] BPT diagram is currently accessible to the Milky Way, our analysis from hereon out is restricted to this diagnostic.

\subsection{Radial Trends}\label{sec:re}
\begin{figure*}[ht!]
\epsscale{1.15}
\plotone{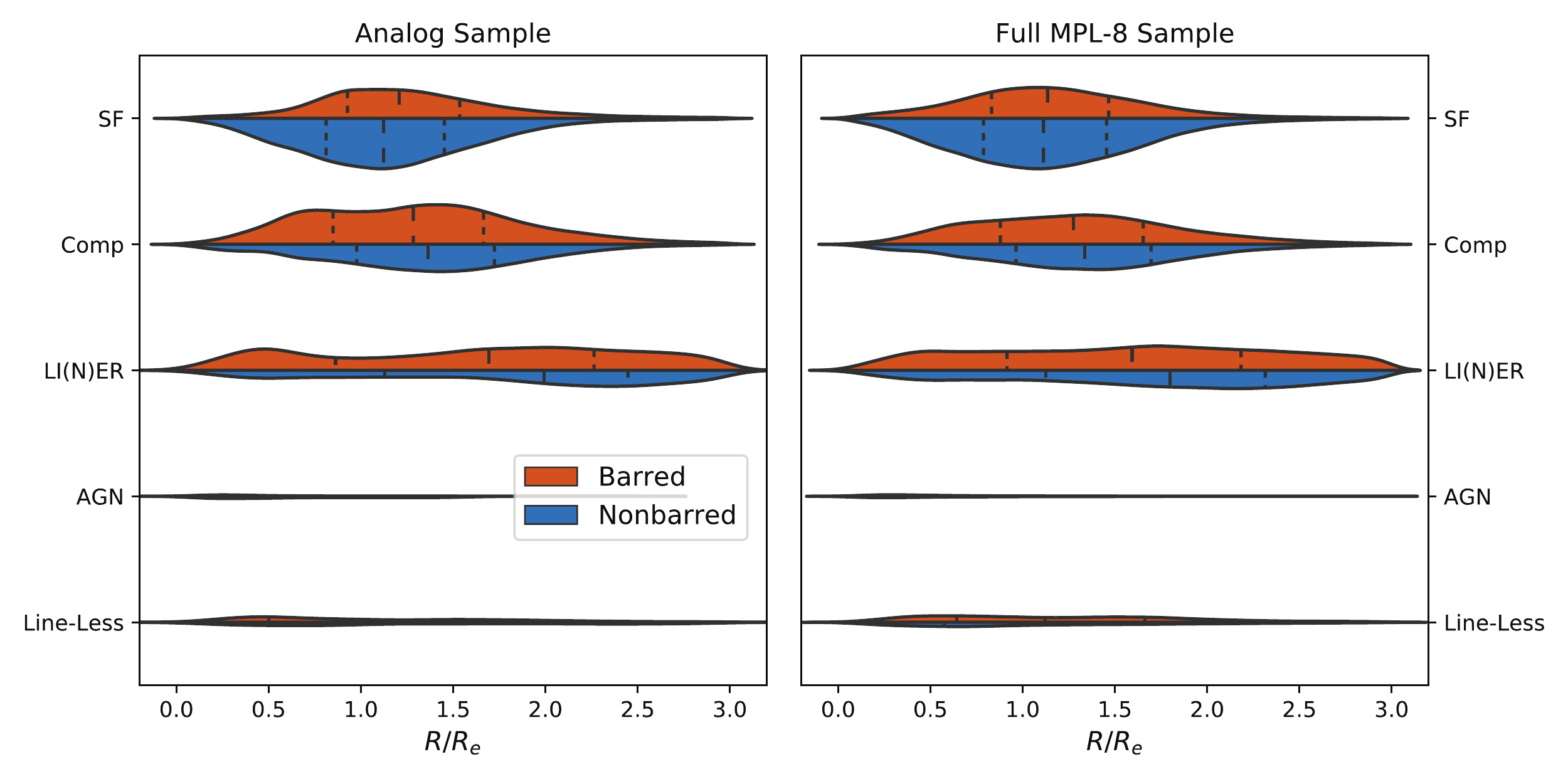}
\caption{Violin style histograms of BPT classifications as a function of $R/R_e$ for barred (orange; upper) and nonbarred (blue; lower) galaxies in the analog (left) and full MPL-8 (right) samples. The interquartile range of each distribution is shown with dashed lines. The shapes of the violin plots are drawn based on the probability density distribution and have their widths normalized based on the number of spaxels in each classification. The relative widths of the violin shapes across different classifications at a given radial bin correspond to the fraction of spaxels within a classification at that bin value. Barred galaxies show a suppression of star formation spectra and an enhancement of composite and LI(N)ER spectra in their inner regions. \label{fig:BPT_Re}}
\end{figure*}

Radial trends are considered for each galaxy sample after normalizing by their effective radii from the NSA catalog \citep{Blanton2011} and correcting for inclination. Figure \ref{fig:BPT_Re} shows the radial distributions of BPT classifications for all four samples of galaxies, revealing differences between the inner regions of barred and nonbarred galaxies. These violin plots show the probability density of the data smoothed by a Gaussian kernel and normalized across all classifications. The ``Line-Less" category shows all spectra that have an \ha\ equivalent width $EW_{\ha} < \unit[1]{\AA}$ and a g-band S/N $> 2$ isolating spaxels within the galaxy but with little to no gas content. Since MaNGA galaxies have more spaxels sampling the outer regions of galaxies than their inner regions, a violin plot of all spaxels in a MaNGA galaxy would have a wedge shape--narrow at small radii and wide at large radii. This style of figures show qualitative differences in the distributions of classifications between samples. At any given radius, the relative widths of distributions show the fraction of spaxels at those classifications. 

In general, star formation is suppressed within the inner $1.5 R_e$ of barred galaxies, where composite and LI(N)ER classifications prevail. Figure \ref{fig:Radial_BPT_KDE} also shows this difference, where the barred galaxies have an increased presence at high [\nii]/\ha\ and [\oiii]/\hb\ line ratios within $R < 0.3 R_e$. In outer regions of galaxies, bars seems to have a lesser effect and star formation dominates the radiation field. LI(N)ER classifications in nonbarred galaxies tend towards larger radii, but do not show a strong difference from barred galaxies at these large radii. AGN and Line-Less spaxels are rare in our samples and are not included in any further analysis.

\begin{figure*}[ht!]

\animategraphics[loop,autoplay,width=7in,controls]{5}{figures_animated/Radial/RadialBPT}{1}{50}

\caption{Slice from an animated figure available online \href{http://www.astronomy.dk/images/MaNGA_BarredAnimations/RadialBPT_KDE_Combined.mp4}{(www.astronomy.dk)} showing Gaussian kernel density estimate BPT-diagrams of barred (orange) and nonbarred (blue) galaxies across different effective radii bins with width $0.3 R_e$. Classification lines are the same as in Figure \ref{fig:BPT_KDE_MW}. Barred galaxies show a flatter distribution across the [\nii]/\ha\ line ratio in their inner regions when compared with nonbarred galaxies and generally show increased central LI(N)ERs. A full description of this 10 second animation can be found in the Appendix Section \ref{sec:fig8}. \label{fig:Radial_BPT_KDE}}
\end{figure*}

\subsection{Total Stellar Mass}\label{sec:mass}
Bars have a clear effect on the ionized gas in galaxies near the mass of the Milky Way, and generally show the same effect across a wide range of stellar masses. However, this may be true for any stellar mass-matched sample, so we compare the radial trends across mass bins in our full MPL-8 samples to consider the importance of the total stellar mass on this effect. Figure \ref{fig:MassBins} shows an animation stepping through mass bins defined such that the sample size of barred galaxies remains constant at $N = 50$. Steps are made in increasing mass to add/remove three galaxies of greater/lower mass from the barred sample. The range in barred galaxy stellar masses in each step sets the galaxies used from the nonbarred sample. Similar to section \ref{sec:re}, radial distributions of BPT classifications and a Gaussian kernel density BPT diagram are shown to compare barred with nonbarred galaxies. 

\begin{figure*}[ht!]

\animategraphics[loop,autoplay,width=7in,controls,poster=41]{5}{figures_animated/MassBin/MassBin}{1}{64}

\caption{Slice from an animated figure available online \href{http://www.astronomy.dk/images/MaNGA_BarredAnimations/MassBinnedBPT.mp4}{(www.astronomy.dk)} showing changes in the BPT classifications of barred (orange; top left) and nonbarred (blue; bottom left) galaxies across different stellar masses. Bins are defined such that the selected barred galaxy sample is fixed to $50$ galaxies and stepped forward to add/remove three galaxies in each step. The nonbarred bins select galaxies within the min/max mass range of the barred galaxies and their sample size varies. The pink and green shaded bars in the stellar mass histograms encompass the mass range used to define the current bin. Violin style distributions of BPT classifications are shown as in Figure \ref{fig:BPT_Re} and a Gaussian kernel density estimate BPT diagram for the binned sample is shown to the right for $R_e < 0.3$. This shows that the differences between gas in barred and nonbarred galaxies extend across a wide range in total stellar mass. A full description of this 13 second animation can be found in the Appendix Section \ref{sec:fig8}.}
\label{fig:MassBins}
\end{figure*}

\begin{figure*}[ht!]
\epsscale{1.15}
\plotone{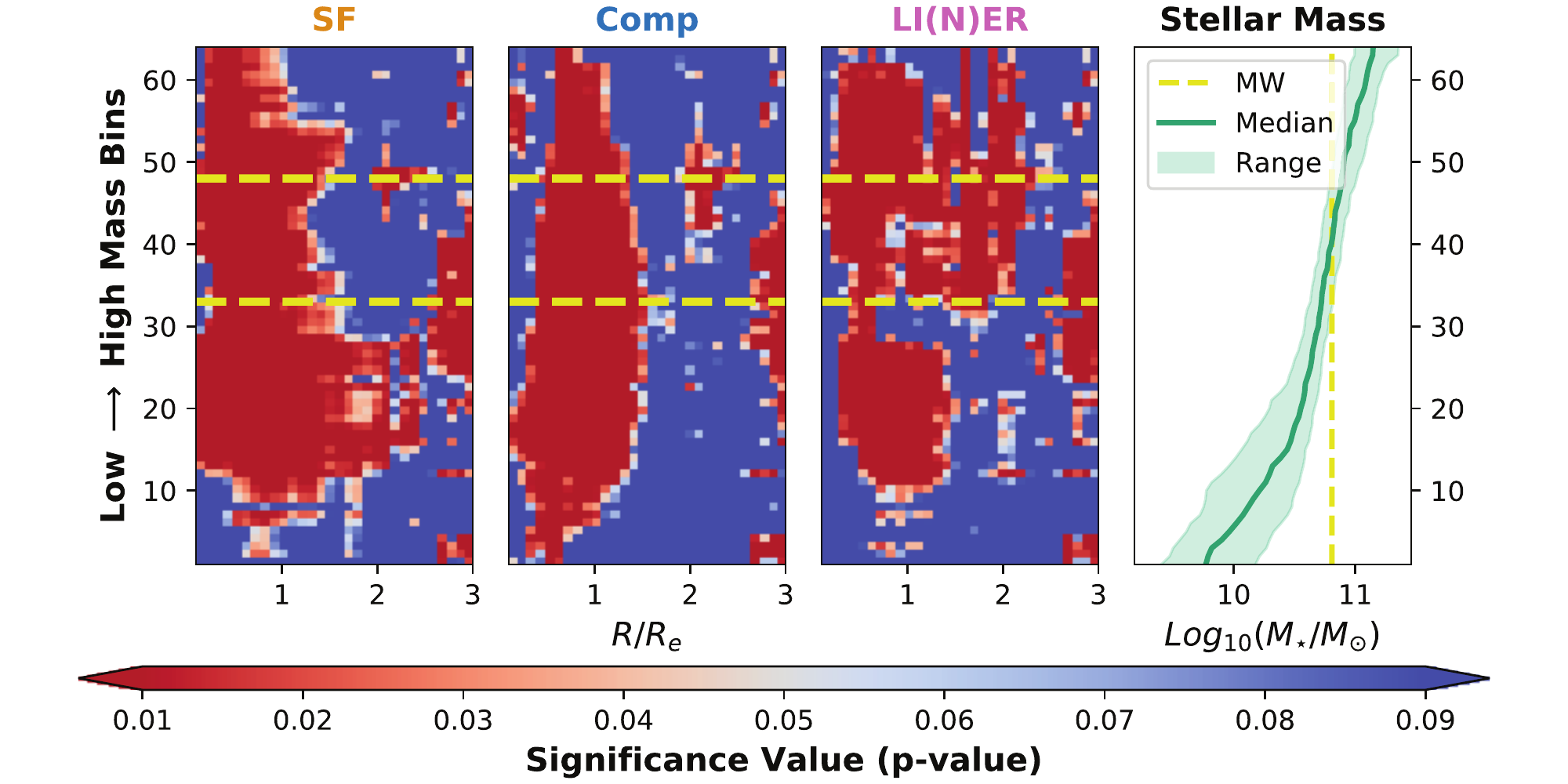}
\caption{Results of the Anderson-Darling statistical test that the underlying distribution of the fraction of spaxels in different BPT classifications come from the same parent distribution across different stellar mass and $0.2 R_e$ radial bins. The transition from red to blue colored pixels is set at the $p = 0.05$ level; deep red points signify that the null hypothesis can be rejected at the $p = 0.01$ significance level and indicate a statistical difference in the fractional classifications. The four panels correspond to the BPT classifications of Star Formation (far left), Composite (mid-left), and LI(N)ER (mid-right). AGN classifications are not considered here since they are an overall rare occurrence in these samples. Stellar mass bins are defined as in Figure \ref{fig:MassBins} such that each mass bin has a fixed barred sample size of $50$ barred galaxies with larger masses situated higher along the y-axis. The median stellar mass, and range of masses considered in each bin is plotted along the far right. The dashed yellow line encloses samples that include a Milky Way stellar mass value of $M = 6.43 \times 10^{10} \textrm{M}_\odot$ \citep{McMillan2011}. We interpret these results to show that the inner regions of barred galaxies, especially around $0.2$ to $1.5 R_e$, have a statistically significant difference from nonbarred galaxies in their ionized gas properties across all but the highest and lowest stellar masses.  \label{fig:Mass_Radius_Stats}}
\end{figure*}

In general, galaxies across all masses show differences in their inner regions when considering the ionized gas. To test this claim, we use the Anderson-Darling statistical test with the null hypothesis that the fraction of BPT classifications as a function of radius are the same for both barred and nonbarred galaxies for all of our mass bins shown in the frames of Figure \ref{fig:MassBins}. Figure \ref{fig:Mass_Radius_Stats} shows the results of this test for three BPT classifications, with each pixel corresponding to the p-value of a test for a given radial bin (x-axis) and mass bin (y-axis). Red points indicate where the null hypothesis can be rejected, indicating a statistical difference between barred and nonbarred galaxies. With the exception of the largest and smallest masses, the inner regions of barred and nonbarred galaxies within $R \lesssim 1.5 R_e$ generally show a statistical difference in their distribution of BPT classifications, while outside of this range, the distributions are indistinguishable. The distribution of LI(N)ERs within the range of Milky Way like masses show a statistical difference across a wider radial range than the other classifications. This may indicate that a loose cut-off mass exists to transition from galaxies in which the bar more significantly affects the ionization conditions of a galaxy. Similarly, star formation classifications show a more wide spread difference at masses in the range of $10.4 \lesssim \log_{10}(\nicefrac{M_\star}{M_\odot}) \lesssim 10.8$. This mass range also corresponds well with signs of inside-out quenching as galaxies transition from star formation to quiescence as shown in \citet[][see their Figure 4]{Belfiore2017}.

\begin{figure}[ht!]
\epsscale{1.15}
\plotone{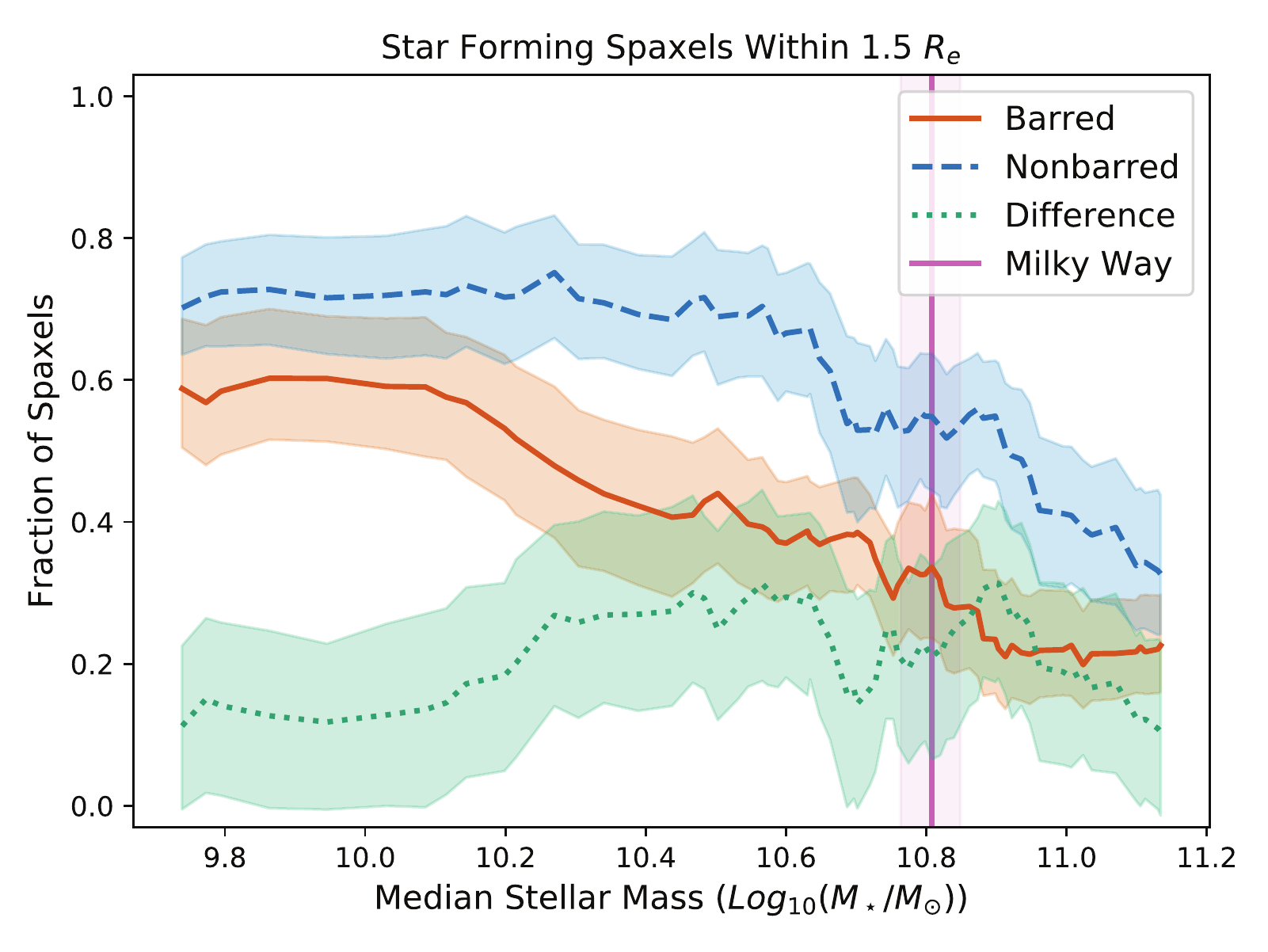}
\caption{Fraction of star formation classification spaxels within the inner $1.5\ R_e$ of galaxies as a function of total stellar mass, where the same mass bins as in Figure \ref{fig:MassBins} and Figure \ref{fig:Mass_Radius_Stats} are used. The median (lines) and $95\%$ confidence interval (shaded regions) are computed using 1000 bootstrap re-samples of size $N = 30$ within each mass bin. The Milky Way stellar mass and $1 \sigma$ error estimate of \citet{McMillan2011} are shown in pink. The nonbarred galaxies (dashed; blue) show a higher fraction of star formation spaxels than barred galaxies (solid; orange), with differences (dotted; green) $\gtrsim 10\%$ across all but the highest masses.  \label{fig:SF_Fractions}}
\end{figure}

To better understand how the presence of star formation differs between the inner parts of barred and nonbarred galaxies, we consider the fraction of spaxels within $1.5\ R_e$ that have star formation BPT classifications in both samples across the same mass bins. The results of this are shown in Figure \ref{fig:SF_Fractions}, where the barred galaxies show a decreased presence of star forming spaxels across all masses. This difference is greatest between $10.3 \lesssim \log_{10}(\nicefrac{M_\star}{M_\odot}) \gtrsim 10.9$, where the nonbarred galaxies show $10\% - 30\%$ greater presence of star forming spaxels than barred galaxies. 

\section{Effect of Bar on Host Galaxy}\label{sec:bar_specific}
Section \ref{sec:stats} establishes an observable and significant impact on the ionized gas in galaxies with a bar compared to those without a bar. Honing in on the barred sample will allow for a more direct measure of how a bar impacts the ISM of its host galaxy. We compare the properties of the ISM interior and exterior to the bar radius and explore differences at a constant radius in the bar and in the inter-bar region. This is accomplished using bar masks from Galaxy Zoo:3D (see Section \ref{sec:obs} for details).  

\begin{figure*}[ht!]

\animategraphics[loop,autoplay,width=7in,controls,poster=41]{5}{figures_animated/BarLength/BarLength}{1}{50}

\caption{Slice from an animated figure available online \href{http://www.astronomy.dk/images/MaNGA_BarredAnimations/BarLength_BPT_Radial_KDE.mp4}{(www.astronomy.dk)} showing changes in the BPT classifications of barred galaxies across different stellar masses (top left) as a function of their bar lengths. The mass bins are chosen to be identical to those in Figure \ref{fig:MassBins} and Figure \ref{fig:Mass_Radius_Stats}. The Gaussian kernel density BPT diagram on the right shows the difference between the classifications in the bar (red), in the inter-bar region (blue), and outside the bar radius (grey). Violin style histograms of the classifications (center) are normalized based on the total number of spaxels at a given radial bin and split into sections inside and outside of the bar radius. The shaded (red) region along the x-axis of this panel is a break in the axis to visually separate the plots at the bar length. The same style of histogram (bottom left) shows the distribution of classifications in the bar (red) and in the inter-bar region (blue) for radii with spaxels found both in the bar and inter-bar region (i.e. $R_{bar,width} < R < R_{bar,length}$). The grey transparent background shows the expected shape if the classifications were uniformly distributed across all spaxels considered. Across a large range of masses, gas inside the bar radius show a decreased presence of star formation and instead is primarily composite and LI(N)ER like. The ionized gas within a bar shows a significantly different radial distribution for star forming and LI(N)ER spaxels compared with ionized gas in the inter-bar region. Composite classifications are smoothly distributed across all radii both inside and outside the bar. A full description of this 13 second animation can be found in the Appendix Section \ref{sec:fig12}. \label{fig:BarLength_BPT}}
\end{figure*}

Figure \ref{fig:BarLength_BPT} shows differences in BPT classifications interior to vs. outside of the bar radius across different masses for $226/240$ barred galaxies in an animated plot. The gas interior to the bar radius shows greatly enhanced [\nii]/\ha\ and [\oiii]/\hb\ line ratios, leading to an increase in LI(N)ER and composite classifications. In the bottom left panel of Figure \ref{fig:BarLength_BPT}, emission from spaxels at radii found both in the bar (red) and inter-bar region (blue) (i.e. $R_{bar,width} < R < R_{bar,length}$) is isolated, similar to \citet{Fraser-McKelvie2019}. The radial distributions of BPT classifications within each of these sub groups vary significantly, with gas in the bar tending towards increased LI(N)ER classifications in the inner portions. Evidence for a star forming ring like structure is seen with the narrow distribution of star formation classifications near the bar radius. Gas in the inter-bar region also seems to be largely composite at larger radii, and LI(N)ER at smaller radii, suggesting that LI(N)ER like gas tends to concentrate in the bar and close to its vicinity. This indicates a steeper gradient in the [\nii]/\ha\ line ratio as a function of radius perpendicular to the bar major axis than parallel with the bar major axis.

\section{Discussion}\label{sec:disc}
LI(N)ERs, and more generally, enhanced [\nii]/\ha\ emission line ratios in galaxies remain complicated and not fully explained since \citet{Burbidge1965} noticed them in the inner regions of galaxies. The previous sections have shown an apparent link between the presence of LI(N)ERs and bars. This apparent link has also been observed in more nearby galaxy samples that find bar regions of galaxies to be largely LI(N)ER \citep{James2015} The bar can be interacting with and setting the conditions of gas in its vicinity in many ways, including increased shear \citep{Renaud2013}, shocks \citep{Athanassoula1982}, enhanced stellar densities \citep{Anders2019}, increased presence of evolved stars, and a non-axisymmetric gravitation potential arising from the stellar structure. 

Inflowing material along the bar can fuel central mass concentrations (such as the CMZ in the Milky Way) or AGN, as observed in other galaxies \citep[e.g.][]{George2019} and in simulations \citep[e.g.][]{Athanassoula1992}. The role of bars in fueling AGN has been largely debated as their role may not be vital \citep{Heckman2014} and there are only slight enhancements of AGN in barred galaxies compared with nonbarred galaxies \citep{Galloway2015}. The increased presence of LI(N)ER like spectra in bars can complicate such studies, as AGN may be wrongly identified if using limited diagnostic methods. 

A proposed source of ionization for LI(N)ERs, and gas with enhanced line ratios in general, are classes of hot evolved stars, such as post-asymptotic giant branch (AGB) stars, subdwarf OB (sdOB) stars, or K and M giants \citep{Belfiore2016, Yan2012, Burbidge1965, James2015}. However, these sources may not emit enough Lyman continuum radiation to account for the ionized gas present as shown locally in the Milky Way \citep{Reynolds1989}, or fail to predict all the observed line ratios \citep{Yan2018}. K and M giants as a source of ionization as described in \citet{Burbidge1963} is distinct in that they provide a source of heating through ejected winds, as opposed to purely photoionzing the gas. This additional heating can help explain the large [\nii]/\ha\ line ratios as an increase in gas temperature (up to $20,000$K). This mechanism can be effective in central parts of galaxies where a large density of these older stars may be present, but would likely not make a significant impact in outer LI(N)ERs \citep[e.g.][]{James2015, Belfiore2016} where photoionization may again dominate. 

If hot evolved stars were primarily responsible for the observed ionization, a high density of stars would be required to produce enough ionizing photons from these relatively low luminosity sources. While at times (e.g. within the bar) the presence of LI(N)ERs are well correlated with large stellar mass surface densities, there are still some regions outside of a bar or generally at areas of low stellar mass surface density that are LI(N)ERs. It may still be possible for a centralized source, or other distant and luminous source to ionize the gas \citep{Domgorgen1994, Wood2004}, especially in a region of the galaxy where the ISM may be evacuated by the bar \citep{Sormani2019}, providing channels for photons to travel large distances. More in-depth modeling of the evolution, radiation field, and distribution of sdOB stars \citep[see][for review]{Heber2016} are necessary to fully constrain the possible ionization sources of LI(N)ERs.

Bars show a strong correlation with the physical conditions of the ionized ISM, with bars generally affecting the gas state within the bar radius and more specifically gas in the bar and inter-bar region (see Figure \ref{fig:BarLength_BPT}). The use of Galaxy Zoo:3D bar masks reveal these differences and demonstrates a clear non-axisymmetric impact of the bar to the ionized gas of galaxies. The presence of LI(N)ER gas in the inter-bar region is especially difficult to ionize from a population of evolved stars since the stellar surface mass density in these regions are generally lower. However, this gas may still feel the effects of an increased population of hot evolved stars within the bar if radiative transfer is able to propagate enough photons radially outside of the bar. In depth $3D$ Monte-Carlo radiative transfer models \citep[see e.g.][]{Wood2010, Vandenbroucke2018} may provide better constraints on the ionization mechanisms at play in the vicinity of a bar. Shocks also may be a viable source of ionization for LI(N)ERs, especially since dust lanes within bars are likely formed via shocks \citep{Athanassoula1992, Englmaier1997, Greve1999, Lindblad1996a, Lindblad1996b}. Better diagnosing for the presence of shocks may be possible with use of the [\sii] $\lambda 6716$ \AA\ emission line in combination with the [\nii] and \ha\ lines and can soon also be tested in the Milky Way using upcoming multi-wavelength WHAM observations of the Tilted Disk \citep[see][]{Krishnarao2020a}. Higher spatial resolution IFU studies of face-on systems, such as the TIMER \citep{Gadotti2019}, PHANGS \citep{Utomo2018, Jiayi2018}, or TYPHOON \citep{Poetrodjojo2019} surveys, may better be able to identify these shock fronts and test for their contributions towards the overall ionization and physical conditions. These nearby galaxies also provide a venue to study the resolved behavior of neutral atomic and molecular gas in HI and CO. While large scale programs to characterize the total HI content of MaNGA galaxies exist \citep[HI-MANGA][]{Masters2019}, only limited and targeted efforts using interferometric observations look to provide spatially resolved information \citep[e.g.][for non-MaNGA galaxies]{Newnham2019}. Understanding the possible role of a galaxy's environment on both bars and their effects on the ISM using existing Value Added Catalogs (VACs) such as the Galaxy Environment for MaNGA Galaxies (GEMA) catalog \citep[e.g.][]{GEMA} is also possible. However, MaNGA does not sample many galaxies in dense clusters, where these environmental effects would be best studied, so such an analysis may be more suitable for other galaxy surveys.

\begin{figure}[ht!]
\epsscale{1.15}
\plotone{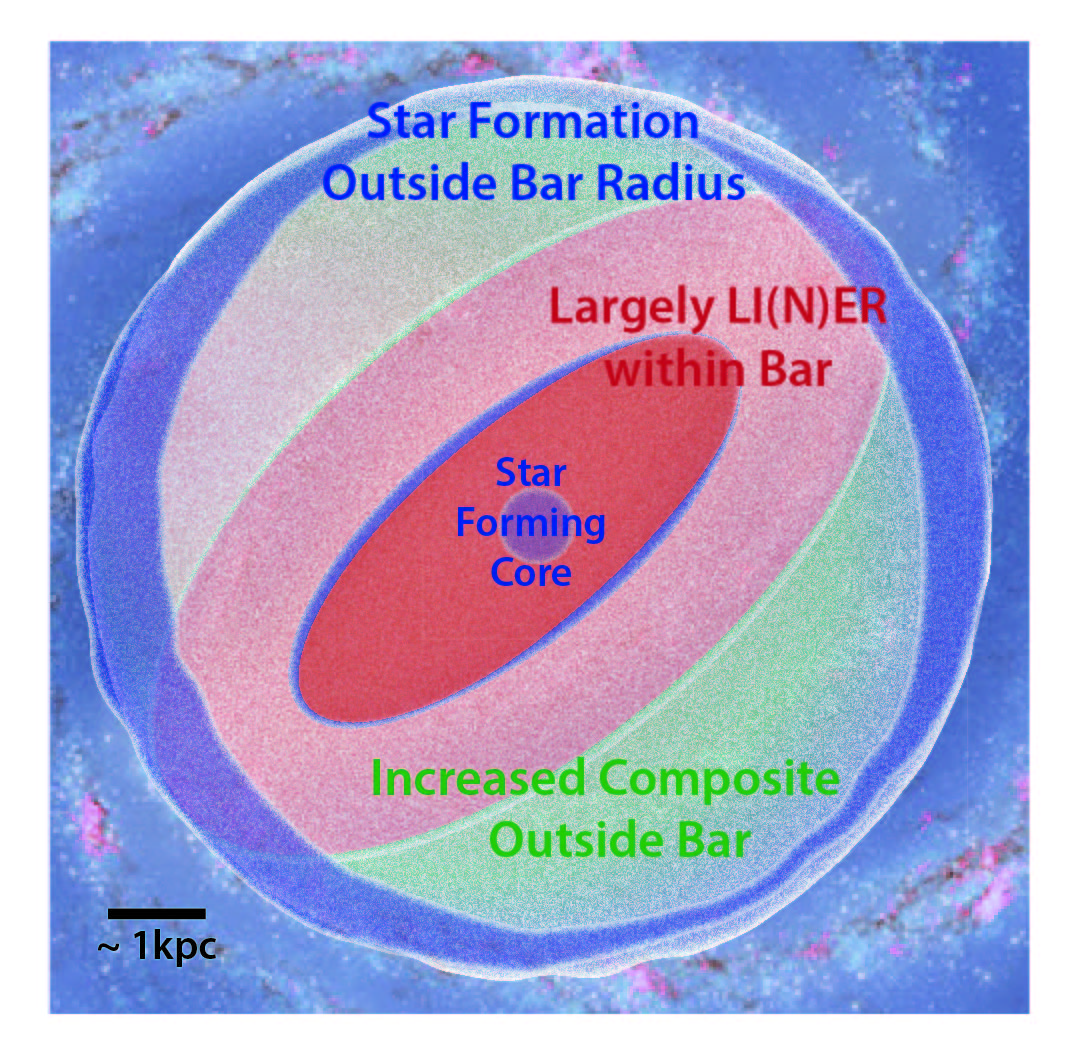}
\caption{A predicted face-on schematic of how the ionized gas properties of the ISM interior to the bar radius may behave in the Milky Way based on Milky Way analogs. Colors correspond to different BPT classifications (star formation: blue, composite: green, LI(N)ER: red). Star formation is seen largely in the central core and some ring like structures near the bar and at/outside the bar radius. Most of the gas in the bar is LI(N)ER like and composite, while most of the gas in the inter-bar region is composite, with some LI(N)ER classifications at smaller radii. \label{fig:CartoonMW}}
\end{figure}

Many of the predictions from simulations \citep[e.g.][]{Athanassoula2002, Renaud2013, Seo2019} and previous observational efforts \citep[e.g.][]{Athanassoula1982, Masters2011, Masters2012} are consistent with the effects seen in this work, such as a suppression of star formation around a bar, with star formation arising in rings at larger radii, or in central cores, like in the Central Molecular Zone \citep[CMZ;][]{Longmore2013}. The Milky Way analogs tend to show consistent behavior with that of the Milky Way, in particular the presence of LI(N)ER classifications within the bar and star formation dominating the radiation field outside the bar. Combining the predictions from simulations, observations in the Milky Way, and analysis of Milky Way analogs in this work, we can speculate on the face-on distribution of ionized gas and dominant ionization mechanisms in the inner Galaxy within the bar radius. 

Figure \ref{fig:CartoonMW} shows a predicted cartoon schematic of how a barred galaxy like the Milky Way may appear when considering the ionized gas conditions. Gas can be funneled in towards the center, where it can condense and form stars in a CMZ-like core. Outside of this region, the gas has large [\nii]/\ha\ line ratios and are largely LI(N)ER, especially within the non-axisymmetric structure of the bar. Along the bars minor axis, but outside of the bar itself, gas transitions from LI(N)ER to composite classifications as the [\nii]/\ha\ line ratio decreases. At around the bar radius, star formation dominates the radiation field in a ring like form and persists throughout the rest of the galaxy as the primary source of ionization.

\section{Conclusions}\label{sec:conclude}

LI(N)ERs have long been seen to be a common occurrence, especially in early type galaxies \citep{Ho1997} and spiral galaxies, with early estimates showing up to $\nicefrac{1}{3}$ of spiral galaxies are LI(N)ERs \citep{Heckman1980}. Here we show that among disk galaxies, barred galaxies show a much greater occurrence of central LI(N)ERs and composite classifications when compared with nonbarred galaxies across all but the smallest and largest masses. This demonstrates an apparent link between the presence of a bar and conditions of ionized gas associated with LI(N)ER like spectra. Many mechanisms can be at play in the environment of a bar, ranging from large shearing motions and shocks commonly observed to suppress star formation. \citet{Burbidge1965} first suggested that gas showing such large ratios of [\nii]/\ha\ may originate from gas that fails to condense, as would be the case for the inner Milky Way and other barred galaxies where the star formation efficiency is low \citep{Longmore2013, Barnes2017}. 

Regardless of the specific mechanisms at play, it is clear that bars have a significant impact on the ionized gas and radiation field in galaxies. Barred galaxies show at least a $10\%$ decrease in the presence of star formation BPT classifications in their inner $1.5\ R_e$ when compared with nonbarred galaxies for all but the highest stellar masses. At around the stellar mass of the Milky Way, this effect is stronger, with a difference of $20\%$ to $30\%$. The presence of a bar may be one of the most important criterion for defining a sample of Milky Way analogs when investigating the ISM, especially in inner regions of galaxies. Additionally, the total stellar mass of a galaxy seems to have relatively little affect on the state of the ionized gas for intermediate masses around that of the Milky Way. This suggests the total stellar mass is not a very important criterion in defining our analog sample. Studying a large sample of face-on barred galaxies has allowed for us to better understand and predict the behavior of the ISM in the Milky Way. Continued studies of the dynamics and physical conditions of gas in barred galaxies and the Milky Way, combined with observations of optical emission, ultraviolet absorption, and sub-mm to radio wavelength studies of the inner Milky Way can provide the highest resolution counterpart to studies of face-on analog samples.

\facilities{
GALEX,
Sloan,
WHAM 
    }

\software{
astropy \citep{astropy,astropy2}, 
matplotlib \citep{mpl}, 
numpy \citep{np}, 
scipy \citep{scipy}, 
seaborn \citep{sns}, 
sdss-marvin \citep{Cherinka2019}, 
pandas \citep{pandas}.  
          }
          
\acknowledgments

DK and CAT acknowledge support from the NSF CAREER Award AST-1554877.

Funding for the Sloan Digital Sky Survey IV has been provided by the Alfred P. Sloan Foundation, the U.S. Department of Energy Office of Science, and the Participating Institutions. SDSS acknowledges support and resources from the Center for High-Performance Computing at the University of Utah. The SDSS web site is www.sdss.org.

SDSS is managed by the Astrophysical Research Consortium for the Participating Institutions of the SDSS Collaboration including the Brazilian Participation Group, the Carnegie Institution for Science, Carnegie Mellon University, the Chilean Participation Group, the French Participation Group, Harvard-Smithsonian Center for Astrophysics, Instituto de Astrof\'{i}sica de Canarias, The Johns Hopkins University, Kavli Institute for the Physics and Mathematics of the Universe (IPMU) / University of Tokyo, the Korean Participation Group, Lawrence Berkeley National Laboratory, Leibniz Institut f\"{u}r Astrophysik Potsdam (AIP), Max-Planck-Institut f\"{u}r Astronomie (MPIA Heidelberg), Max-Planck-Institut f\"{u}r Astrophysik (MPA Garching), Max-Planck-Institut f\"{u}r Extraterrestrische Physik (MPE), National Astronomical Observatories of China, New Mexico State University, New York University, University of Notre Dame, Observat\'{o}rio Nacional / MCTI, The Ohio State University, Pennsylvania State University, Shanghai Astronomical Observatory, United Kingdom Participation Group, Universidad Nacional Aut\'{o}noma de M\'{e}xico, University of Arizona, University of Colorado Boulder, University of Oxford, University of Portsmouth, University of Utah, University of Virginia, University of Washington, University of Wisconsin, Vanderbilt University, and Yale University.

 \newcommand{\noop}[1]{}

\bibliographystyle{aasjournal}

\appendix

\section{Galaxy Zoo:3D - Bar Length Validation}\label{sec:bar_valid}

Bar lengths measured using Galaxy Zoo:3D bar masks are validated with an independent measure of the bar length. Bar lengths are measured following the methodology employed in \citet{Kraljic2012}. This method decomposes stellar surface density profiles inferred from white light images into azimuthal Fourier components. The signature of a bar is seen in the even ($m=2$) Fourier component within the bar region. Bars can be identified when the $\Phi_{2}(r)$ phase is constant with radius. This method requires a reliable estimate of the bulge-to-total ratio (BTR) of the host galaxy to correctly guess the bar starting position. BTR r-band measurements from two-component decompositions of \citet{Simard2011} are used for these initial guesses. The presence of an additional bar component can bias the amount of light attributed to the bulge, and hence give a BTR that is too large for a galaxy. When this happens, the starting position of the bar-finding code is too large in radius, and the bar may not be correctly identified. Out of the $240$ barred galaxies in our full MPL-8 sample (described in Section \ref{sec:sample}), $206$ returned reasonable measurements.

\begin{figure}[ht!]
\epsscale{1.15}
\plotone{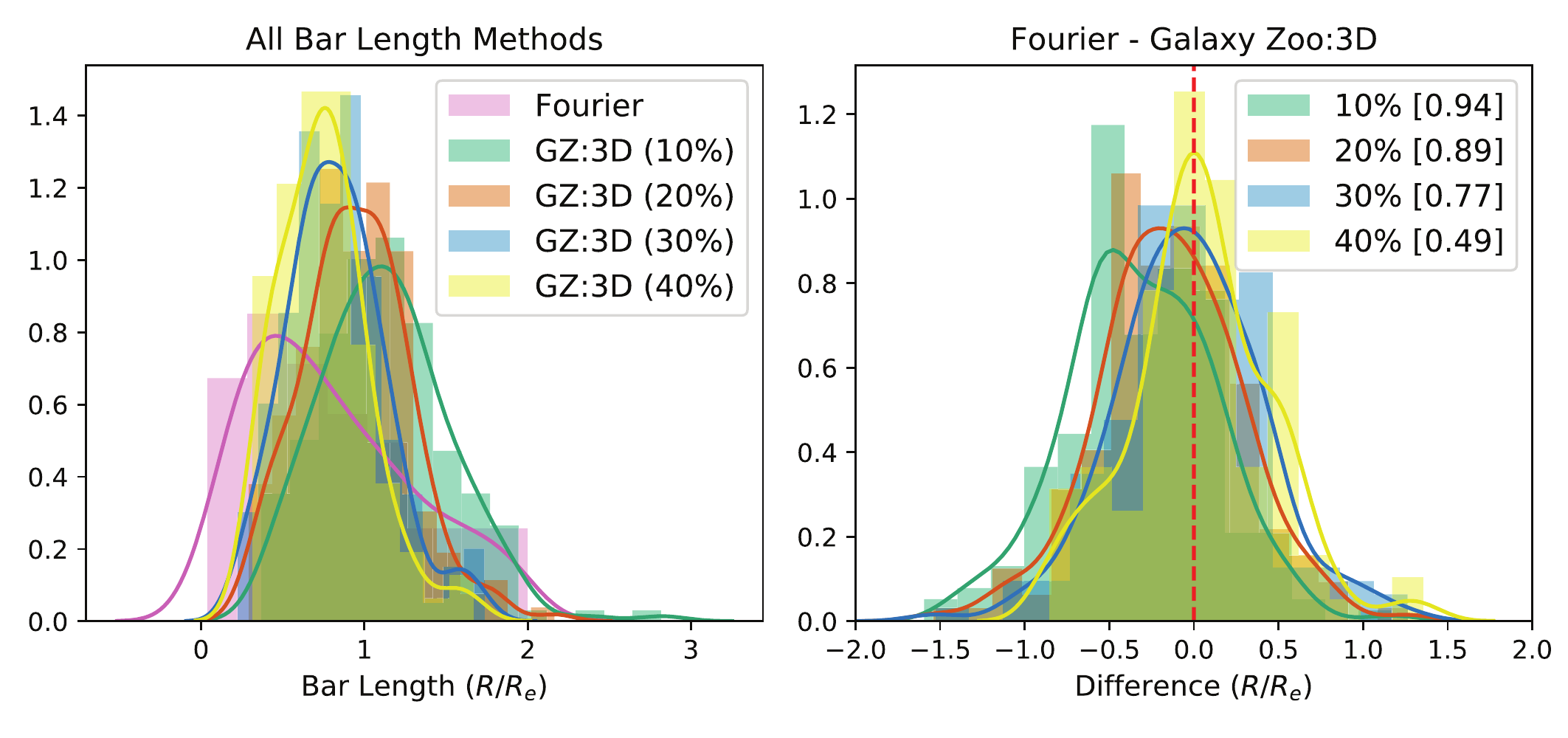}
\caption{Histogram of bar length measurements from the Fourier method employed in \citet[][pink]{Kraljic2012}, and from bar masks using Galaxy Zoo:3D with vote thresholds of $10\%$ (green), $20\%$ (orange), $30\%$ (blue), and $40\%$ (yellow) on the left panel. The fraction of galaxies from the barred MPL-8 sample that have valid bar masks with each threshold is shown in square brackets. Increasing the vote threshold parameter increases the agreement between bar length measurements from both methods, but greatly decreases the number of galaxies that return valid bar masks. The right panel shows histograms of the differences in bar length measurements between the Fourier method and Galaxy Zoo:3D, color coded by vote thresholds. \label{fig:BarLength_Validation}}
\end{figure}

Figure \ref{fig:BarLength_Validation} shows a histogram of the measured bar lengths using the Fourier method and various vote thresholds for Galaxy Zoo:3D. The vote threshold parameter is the percent of votes that must be present to flag a spaxel as part of the bar mask. Increasing this threshold selects fewer spaxels in a galaxy, but can help to ensure the bar mask is isolated to spaxels with a high confidence of being in the bar. Increasing this threshold also reduces total number of galaxies that have valid bar masks. On the lower panel of Figure \ref{fig:BarLength_Validation}, the difference between the Fourier bar length measurement and bar mask measurements for different vote thresholds are shown. Increasing the threshold improves agreement between both methods, suggesting that too many spaxels are selected for the bar mask with low thresholds. However, the completeness of selecting barred galaxies greatly diminishes with larger threshold values. This is seen by the fraction of galaxies that return a valid bar mask for each threshold, shown in square brackets on the Figure \ref{fig:BarLength_Validation} legend. To balance these effects, we use a vote threshold of $20\%$ in this work. Previous work using earlier versions of Galaxy Zoo:3D also used a $20\%$ vote threshold \citep[e.g.][]{Fraser-McKelvie2019}.

\section{Animated Figure Descriptions}
Full descriptions of the contents of animated figures used in this work are available below. 

\subsection{Figure 8}\label{sec:fig8}
The animation in Figure \ref{fig:Radial_BPT_KDE} runs for 10 seconds and steps through radial bins, starting at $0.0 < \nicefrac{R}{R_e} < 0.3$ and ending at $2.9 < \nicefrac{R}{R_e} < 3.1$. The distributions of both barred and nonbarred galaxies in both the analog and full MPL-8 samples have the largest dispersion in the inner most radial bin (corresponding to the still image in the print version). At larger radii, the presence of LI(N)ER classifications decreases, star forming classifications increase, and in general, the [\nii]/\ha\ line ratio decreases. At large radii, both distributions move towards the upper left corner of the diagram within the star forming and composite classifications. The greatest difference between the barred and nonbarred samples are seen within $\sim 1 R_e$, with barred galaxies showing a significantly larger width along both axes, compared to a more strongly peaked distribution for nonbarred galaxies.

\subsection{Figure 9}\label{sec:fig9}
The animation in Figure \ref{fig:MassBins} runs for 13 seconds and steps through total stellar mass bins starting at $9.29 < \log_{10}(M/M_{\odot}) < 10.13$ and ending at $10.99 < \log_{10}(M/M_{\odot}) < 11.35$. As the animation progresses, the stellar mass distributions in the left panel remain the same and the pink and orange bars slide from the left to right to encompass the current range of masses being considered in the middle and right panels. The violin distributions in the middle panel show that star forming classifications are equally dominant in the inner parts of barred and nonbarred galaxies at the lowest masses, while their presence in barred galaxies decreases at larger stellar masses. Composite classifications are uncommon in the lowest masses and gradually increase in their presence at larger masses. LI(N)ER classifications show a larger presence in barred galaxies, especially within $1.5 R_e$ at lower masses, and continue to show this trend at almost all mass bins but the very largest. On the right panel, the BPT diagram is concentrated primarily to the star forming classification at the lowest masses and expands to encompass composites and LI(N)ERs at larger masses.

\subsection{Figure 12}\label{sec:fig12}
The animation in Figure \ref{fig:BarLength_BPT} runs for 13 seconds, and steps through total stellar mass bins as in Figure \ref{fig:MassBins}. The distribution of total stellar mass in the upper left panel remains the same and the pink bar slides from the left to right to encompass the current range of masses being considered in the other panels. In the lower left panel star forming classifications initially dominate at the lowest masses and decreases at larger masses, becoming relatively rare in both the bar and inter-bar regions. Composite classifications start very low and increase to dominate the inter-bar region at all radii at larger masses. LI(N)ER classifications also start very low and then rise at larger masses to dominate the inner regions of both the bar and inter-bar region. The middle panel shows the same overall trends, but now considering all spaxels equally within the bar radius and outside the bar radius. At the lowest masses, the animation begins to show the star forming classification dominates at all radii and gradually decreases, especially within the bar radius, at larger masses. Composite classifications gradually rise in prominence, especially near and beyond the bar radius. LI(N)ER classifications rise to dominate within the bar radius, and are also somewhat present at larger radii at larger masses. The right panel shows that the bar, inter-bar region, and the region outside the bar radius are relatively similar at the lowest masses. As the animation progresses, the region outside the bar radius remains relatively in place while the bar and inter-bar distributions spread out along both axes. The bar region undergoes this spread faster and develops a peak in its distribution at larger values of the line ratios.

\end{document}